\documentclass{amsart}

\usepackage{graphicx}
\usepackage{amssymb}
\newcommand{\given}{\,|\,}

\DeclareSymbolFont{bbold}{U}{bbold}{m}{n}
\DeclareSymbolFontAlphabet{\mathbbold}{bbold}
\newcommand{\ind}{\mathbbold{1}}

\usepackage{natbib}

\usepackage{lineno}
\linenumbers

\numberwithin{equation}{section}
\numberwithin{figure}{section}

\allowdisplaybreaks

\begin{document}

\title[Wright-Fisher bridges]{Analysis and rejection sampling of Wright-Fisher diffusion bridges}

\author{Joshua G. Schraiber}

\address{Department of Integrative Biology\\
 University of California\\
 3060 Valley Life Sciences Bldg \#3140\\
 Berkeley, CA 94720-3140\\
 U.S.A.}

\email{jgschraiber@berkeley.edu}
\thanks{JGS supported in part by NIH NRSA trainee appointment grant 
T32 HG 00047 and by NIH grant R01-GM40282}

\author{Robert C. Griffiths}

\address{Department of Statistics\\ 
1 South Parks Road\\ 
Oxford OX1 3TG\\ 
U.K.} 

\email{griff@stats.ox.ac.uk}
\thanks{RCG supported by Miller Institute for Basic Research in Science, 
University of California at Berkeley}

\author{Steven N. Evans}

\address{Department of Statistics\\
         University  of California\\
         367 Evans Hall \#3860\\
         Berkeley, CA 94720-3860 \\
         U.S.A.}

\email{evans@stat.berkeley.edu}
\thanks{SNE supported in part by NSF grant DMS-0907630}

\date{\today}                                           

\begin{abstract}
We investigate the properties
of a Wright-Fisher diffusion process started
from frequency $x$ at time $0$ and conditioned to be at
frequency $y$ at time $T$.  Such a process is called a
bridge. Bridges arise naturally in the
analysis of selection acting on standing variation and
in the inference of selection from allele frequency time series.
We establish a number of results about the distribution of neutral
Wright-Fisher bridges and develop a novel rejection sampling scheme
for bridges under selection that we use to study their behavior.
\end{abstract}

\maketitle

\section{Introduction}

The Wright-Fisher Markov chain is of central importance in population genetics 
and has contributed greatly to the understanding of the patterns of genetic 
variation seen in natural populations. Much recent work has focused on 
developing sampling theory for neutral sites linked to sites under selection 
\citep{Smith1974, Kaplan1989, Nielsen2005, Etheridge2006}. Typically, the site 
under selection is assumed to have dynamics governed by the diffusion process 
limit of the Wright-Fisher chain, in which case the genealogy of linked 
neutral sites can be constructed using the framework of \citet{Hudson1988}. 
However, due to the complicated nature of this model, analytical theory is 
necessarily approximate and the main focus is on simulation methods. 
In particular, a number of simulation programs, including mbs \citep{Teshima2009}
and msms \citep{Ewing2010} have recently appeared to help facilitate 
the simulation of neutral genealogies linked to sites undergoing a Wright-Fisher diffusion with selection.

Simulations of Wright-Fisher paths under selection can be easily carried out using standard techniques for simulating diffusions. Frequently, however, it is necessary to simulate a Wright-Fisher path conditioned on some particular outcome. For example, to simulate the path of an allele under selection that is currently at frequency $x$, a time-reversal argument shows that it is possible to simulate a path starting at $x$ conditioned to hit $0$ eventually \citep{Maruyama1974}. However, more complicated scenarios, including the action of natural selection on standing genetic variation, require more 
elaborate simulation methods \citep{peter2012distinguishing}. 

The stochastic process describing an allele that starts at frequency $x$ at time $0$ and is conditioned to end at frequency $y$ at time $T$ 
is called a bridge between $x$ and $y$ in time $T$ or a bridge between $x$ and $y$ over the time
interval $[0,T]$. Wright-Fisher diffusion bridges appear naturally 
in the study of selection acting on standing variation because it is necessary to know the path taken by an allele at current frequency
$y$ that fell under the influence of natural selection at a time $T$ generations in the past when it was segregating neutrally at frequency $x$. Wright-Fisher diffusion bridges are also of interest for their application to inference of selection from allele frequency time series \citep{Bollback2008, malaspinas2012estimating, mathieson2013estimating, feder2013identifying}. In particular, analysis of bridges can help determine 
the extent to which more signal is gained by adding further 
intermediate time points.

In addition to their applied interest, there are interesting theoretical questions surrounding Wright-Fisher diffusion bridges. For alleles conditioned to eventually fix, \citet{Maruyama1974} showed that the 
distribution of the trajectory does not depend on the sign of the selection coefficient; that is, both positively and negatively selected alleles 
with the same absolute value of the selection coefficient
exhibit the same dynamics conditioned on eventual fixation. 
It is natural to inquire whether
 the analogous result holds for a bridge between any two interior points. Moreover, the degree to which a Wright-Fisher bridge with selection will differ from a Wright-Fisher bridge under neutrality is not known (in connection
 with this question, we recall the well-known fact that the distribution
 of a bridge for a Brownian motion with drift does not depend on
 the drift parameter, and so it is conceivable that 
 the presence of selection has
 little or no effect on the behavior of Wright-Fisher bridges).  
Lastly, the characteristics of the sample paths of the frequency
of alleles destined to be lost in a fixed amount of time are not only interesting theoretically 
but may also have applications to geographically structured populations \citep{slatkin2012serial}.

Here we investigate various features of Wright-Fisher diffusion bridges. 
The paper is structured as follows. First, we establish analytical results for neutral Wright-Fisher bridges. Then, we derive a novel rejection sampler for Wright-Fisher bridges with selection and use it to study
the properties of such processes. For example, we
estimate the distribution of the maximum of a bridge from 
$0$ to $0$ under selection and investigate how this
distribution depends on the strength of selection.

\section{Background}

A Wright-Fisher diffusion with genic selection is a diffusion process 
$\{X_t, \, t\geq 0\}$  
with state space $[0,1]$ and infinitesimal generator
\begin{equation}
\mathcal{L} = \gamma x(1-x)\frac{\partial}{\partial x} 
+ \frac{1}{2}x(1-x)\frac{\partial ^2} {\partial x^2}.
\label{gen}
\end{equation}
When $\gamma = 0$, the diffusion is said to be neutral; otherwise, the drift term captures
the strength and direction of natural selection.

The corresponding Wright-Fisher diffusion bridge, $\{X_t^{x,z,[0,T]}, \, 0\leq t \leq T\}$ is the stochastic process that results from conditioning the Wright-Fisher diffusion to start with value $x$ at time $0$ and end with value $z$ at time $T$. Denote by $f(x,y;t)$ the 
transition density of the diffusion corresponding to \eqref{gen}.
 By the Markov property of the Wright-Fisher diffusion,
the bridge is a time-inhomogeneous diffusion and
the transition density for the bridge
going from state $u$ at time $s$ to state $v$ at time $t$ is 
\begin{equation}
f_{x,z,[0,T]}(u,v; s,t) = \frac{f(u,v;t-s)f(v,z;T-t)}{f(u,z;T-s)}.
\label{bridge}
\end{equation}
The time-inhomogeneous infinitesimal generator of the bridge acting on a test 
function $g$ at time $s$ is
\begin{equation}
\label{bridgegen}
\begin{split}
\mathcal{L}_{x,z,[0,T];s} g(u) &= \lim_{t \downarrow s} \frac{\mathbb{E}[g(X_t) \given X_0 = x, X_s = u, X_T = z] - g(u)}{t-s}  \\
& = u(1-u)\left(\gamma + \frac{\partial}{\partial u}\log f(u,z;T-s)\right)\frac{\partial g}{\partial u} (u) \\
& \quad + \frac{1}{2}u(1-u)\frac{\partial^2 g}{\partial u^2}(u). \\
\end{split}
\end{equation}

An obvious method for simulating a Wright-Fisher bridge would be
to simulate the stochastic differential equation (SDE)
corresponding to this infinitesimal generator.  There are two obstacles
to this approach.  Firstly, analytic expressions for the transition density
$f$ are only known for the neutral case, and even there they are in
the form of infinite series.  Secondly,
note that the first order coefficient in the 
infinitesimal generator becomes increasing singular
as $s \uparrow T$;  consequently, an attempt to simulate
the bridge by simulating the SDE would be quite unstable because
the drift term in the SDE would explode at times close to the
terminal time $T$.  It is because this naive approach is infeasible
that we need to consider the more sophisticated simulation methods
explored in this paper.

In addition to conditioning the process to obtain a particular value at a particular time, it is possible to condition a process's long term behavior. The transition densities of the conditioned process, $f_h(x,y;t)$ are related to to the transition densities of the unconditioned process by the usual Doob $h$-transform formula,
\[
f_h(x,y;t) := h(x)^{-1}f(x,y;t)h(y).
\]
The $h$-transformed process has infinitesimal generator
\begin{equation}
\mathcal{L}^h := x(1-x)\left(\gamma + \frac{h'(x)}{h(x)} \right)\frac{\partial}{\partial x} + \frac{x(1-x)}{2}\frac{\partial^2}{\partial x^2}.
\label{htrans}
\end{equation}
Note that the finite dimensional marginal distribution at times $0 \leq t_1 \leq \ldots \leq t_n \leq T$ of the Wright-Fisher diffusion bridge starting at $x$ at time $0$ and ending at $y$ at time $T$ has density
\[
\frac{f(x,v_1;t_1)f(v_1,v_2;t_2-t_2)\cdots f(v_n,y;T-t_n)}{f(x,y;T)}
\]
whereas the analogous density for the corresponding bridge of the $h$-transformed process is
\[
\begin{split}
&\frac{h(x)^{-1}f(x,v_1;t_1)h(v_1)h(v_1)^{-1}f(v_1,v_2;t_2-t_1)h(v_2)\cdots h(v_n)^{-1} f(v_n,y;T-t_n)h(y)}{h(x)^{-1}f(x,y;T)h(y)} \\
& \quad = \frac{f(x,v_1;t_1)f(v_1,v_2;t_2-t_1)\cdots f(v_n,y; T-t_n)}{f(x,y;T)}.
\end{split}
\]
Thus, the the bridges for the two processes have the same distribution. 

Typical $h$-transforms include the conditioning a process to eventually
hit a particular value, and for the sake of future reference we
recall from standard diffusion theory \citep{MR1780932} that
the probability that the Wright-Fisher diffusion
started from $x$ eventually hits $y$ is
\begin{equation}
p_{xy} = 
\begin{cases}
\frac{S(x)-S(0)}{S(y)-S(0)}, & \text{if } y > x, \\
\frac{S(1)-S(y)}{S(1)-S(x)}, & \text{if } y < x,
\end{cases}
\end{equation}
where $S$ is the scale function given by
\[
S(x) = 
\begin{cases} 
\frac{1-e^{-2\gamma x}}{1-e^{-2\gamma}}, & \text{if $\gamma \ne 0$}, \\
x, & \text{if $\gamma = 0$}.
\end{cases}
\]
Thus, 
\begin{equation}
p_{xy} =
\begin{cases}
\frac{1-e^{-2\gamma x}}{1-e^{-2\gamma y}}, & \text{if } y > x, \\
\frac{e^{-2\gamma y} - e^{-2\gamma }}{e^{-2\gamma x}-e^{-2\gamma}}, & \text{if } y < x,
\end{cases}
\label{fp}
\end{equation}
when $\gamma \ne 0$ and
\begin{equation}
p_{xy} =
\begin{cases}
\frac{x}{y}, & \text{if } y > x, \\
\frac{1-y}{1-x}, & \text{if } y < x.
\end{cases}
\label{fp_neutral}
\end{equation}

\section{Analytic theory for neutral bridges}

\subsection{Transition densities for the neutral Wright-Fisher diffusion}

When there is no natural selection (i.e., $\gamma = 0$),
the transition densities of the Wright-Fisher diffusion can be expressed
\begin{equation}
f(x,y;t) = \sum_{l=2}^\infty q_l(t) \sum_{k=1}^{l-1}\binom{l}{k}x^k(1-x)^{l-k}\mathcal{B}(y;k,l-k),
\label{transition}
\end{equation}
where the $q_l(t)$ are the transition functions of a death process starting at infinity with death rate $\frac{1}{2}n(n-1)$ when $n$ individuals are left
alive and $\mathcal{B}(\cdot;\alpha,\beta)$ is the density of the Beta distribution with parameters $\alpha$ and $\beta$ \citep{MR1235429}.
That is, $q_l(t)$ is the probability that a Kingman coalescent tree
with infinitely many leaves at time $0$ has $l$ lineages present
$t$ units of time in the past.  
In the Appendix we present a related pair of eigenfunction expansions 
of the transition density.

Let $\{T_{j}\}_{j=1}^\infty$ be a sequence of
independent exponential random variables with rates $\{j(j-1)/2\}_{j=1}^\infty$.  We think of 
$T_{j}$ as the length of time in a Kingman coalescent tree 
when $j$ lineages are present.  Thus, $\sum_{j=l}^\infty T_j$
is the time to $l-1$ lineages being present.
Write $h_l(t)$ for the density of this sum.
The Laplace transform of $h_l$ is
\begin{eqnarray}
\phi_l(\lambda) &=& \int_0^\infty e^{-\lambda t}h_l(t)dt \nonumber \\
&=& \prod_{j=l}^\infty \left(1 + \frac{2\lambda}{j(j-1)} \right)^{-1}.
\end{eqnarray} 
Because 
\[
h_l(t) = \frac{1}{2}l(l-1)q_l(t), \quad t > 0,
\]
we see that
\begin{equation}
\int_0^\infty e^{-\lambda t}q_l(t)dt = \frac{2}{l(l-1)}\phi_l(\lambda), \quad l>0.
\end{equation}
Thus, the Laplace transform of $f(x,y;\cdot)$ is
\begin{equation}
f^*(x,y; \lambda) = \sum_{l=2}^\infty \frac{2}{l(l-1)}\phi_l(\lambda)\sum_{k=1}^{l-1}\binom{l}{k}x^k(1-x)^{l-k}\mathcal{B}(y; k, l-k).
\label{laplace}
\end{equation}

To construct bridges with $0$ as their initial or final points, 
we need to consider the behavior of the transition density 
$f(x,y;t)$ as $x \downarrow 0$. Discarding terms that
are $O(x^2)$, \eqref{laplace} is asymptotic to
\begin{equation}
2x\sum_{l=2}^\infty(1-y)^{l-2}\phi_l(\lambda).
\label{asymp}
\end{equation}
Note that 
\begin{equation}
\sum_{l=2}^\infty y(1-y)^{l-2}\phi_l(\lambda)
\end{equation}
is the Laplace transform of the density of 
\begin{equation}
\sum_{l=N}^\infty T_l,
\end{equation}
where $N-2$ is distributed as the number of failures
before the first success in a sequence of i.i.d. Bernoulli
trials with success probability $y$.

\subsection{Bridge from $0$ to $0$ over $[0,T]$} 

For $x,y \notin \{0,1\}$, it follows from \eqref{bridge}  that 
the density of $X_t$ given that $X_0 = x$ and $X_T = z$ is
\begin{eqnarray}
f_{x,z,[0,T]}(y; t) &=& \frac{f(x,y;t)f(y,z;T-t)}{f(x,z;T)} \nonumber \\
&=& \frac{f(x,y;t)f(z,y;T-t)y(1-y)}{f(x,z;T)z(1-z)} \nonumber \\
&=& \frac{x^{-1}f(x,y;t)z^{-1}f(z,y,T-t)y(1-y)}{x^{-1}f(x,z;T)(1-z)}.
\label{condition}
\end{eqnarray}
In the second line of \eqref{condition} we used reversibility (before hitting 0 or 1) with respect to the speed measure $z^{-1}(1-z)^{-1}$. 
From \eqref{laplace} we know the asymptotic form of \eqref{condition}. The limit of 
\[
x^{-1}f(x,z;T)
\]
as $x\downarrow 0$ is
\begin{equation}
2\sum_{l=2}^\infty (1-z)^{l-2}h_l(T)
\label{xtozero}.
\end{equation}
If $z \downarrow 0$ as well, then the limit is
\begin{equation}
2\sum_{l=2}^\infty h_l(t).
\end{equation}
Therefore,
\begin{equation}
\begin{split}
& f_{0,0,[0,T]}(y;t) \\
& \quad =\frac{2y(1-y)\sum_{k=2}^\infty(1-y)^{k-2}h_k(t)
\times\sum_{l=2}^\infty(1-y)^{l-2}h_l(T-t)}{\sum_{m=2}^\infty h_m(T)}. \\
\label{0to0}
\end{split}
\end{equation}
The density $h_l$ is given by
\begin{equation}
h_l(t) = \frac{1}{2}l(l-1)\sum_{j=l}^\infty 
e^{-\frac{j(j-1)}{2}t}(-1)^{j-l}\frac{(2j-1)l_{(j-1)}}{l!(j-l)!},
\label{hl}
\end{equation}
where $a_{(b)} := a(a+1)\cdots(a+b-1)$. 
In addition, an eigenfunction expansion of the transition density 
in the Appendix shows that
\begin{equation}
2\sum_{l=2}^\infty h_l(t) = \sum_{n=2}^\infty e^{-\frac{1}{2}n(n-1)t}(2n-1)n(n-1).
\end{equation}

It is clear from the above that the random variable
$X_t^{0,0,[0,T]}$ has the same
distribution as $X_{T-t}^{0,0,[0,T]}$ for $0 \le t \le T$,
and an elaboration of
this argument using \eqref{bridge}
to compute the finite dimensional distributions of  the process
$X^{0,0,[0,T]}$ shows the following invariance under
time-reversal 
\[
\{X_t^{0,0,[0,T]}, \; 0 \leq t \leq T\} 
\overset{\mathcal{D}}{=} 
\{X_{T-t}^{0,0,[0,T]}, \; 0 \leq t \leq T\}, 
\]
where $\overset{\mathcal{D}}{=}$ denotes equality in distribution. 

As $T\rightarrow \infty$, the density of $X_t^{0,0,[0,T]}$  
for a fixed $t>0$ converges to
\begin{equation}
2y(1-y)e^t\sum_{k=2}^\infty(1-y)^{k-2}h_k(t).
\end{equation}

By a similar calculation, 
we find that, centering around $T/2$, 
the limiting density of $X_{T/2+t}$ for $-T/2 < t  < T/2$ fixed is just
$6y(1-y)$,
independent of $t$.

Moreover, from \eqref{bridge} we see that the transition densities
of $X_t^{0,0,[0,T]}$ satisfy
\begin{equation}
\begin{split}
f_{0,0,[0,T]}(u,v; s,t) 
& = \lim_{z \downarrow 0} \frac{f(u,v;t-s)f(v,z;T-t)}{f(u,z;T-s)} \\
& = \lim_{z \downarrow 0} \frac{f(u,v;t-s)f(v,z;T-t)z(1-z) }{f(u,z;T-s)z(1-z)} \\
& = \lim_{z \downarrow 0} \frac{f(u,v;t-s)f(z,v;T-t)v(1-v) }{f(z,u;T-s)u(1-u)} \\
& = f(u,v;t-s) \frac{\sum_{l=2}^\infty (1-v)^{l-2}h_l(T-t) v(1-v)}
{\sum_{l=2}^\infty (1-v)^{l-2}h_l(T-s) u(1-u)}. \\
\end{split}
\end{equation}
For fixed $0 < s < t$, this transition density converges to
\begin{equation}
\lim_{T \to \infty}
f_{0,0,[0,T]}(u,v; s,t) = e^{t-s} u^{-1} (1-u)^{-1} f(u,v;t-s) v(1-v),
\label{noabsorb_transition}
\end{equation}
the transition density of the neutral Wright-Fisher diffusion
conditioned on non-absorption, a process with infinitesimal generator
\begin{equation}
(1 - 2y) \frac{\partial}{\partial y} + \frac{1}{2} y (1-y) \frac{\partial^2}{\partial y^2}.
\label{noabsorb_generator}
\end{equation}
For fixed $-\infty < s < t < \infty$,
the transition density $f_{0,0,[0,T]}(u,v; T/2+s,T/2+t)$ converges 
as $T \to \infty$ to the same limit, 
and so the finite-dimensional distributions of
the process $\{X_{T/2+t}^{0,0,[0,T]}, \; -T/2<t<T/2\}$ converge
to those of the stationary Markov process indexed by the whole
real line that is obtained by taking the neutral Wright-Fisher diffusion
conditioned on non-absorption in equilibrium.

\subsection{Bridge from $x$ to $0$ over $[0,T]$}

The density of $X_t$ given that $X_0 = x$ and $X_T = 0$ is
\begin{equation}
f_{x,0,[0,T]}(y;t) 
= 
f(x,y;t)\frac{\sum_{l=2}^\infty y(1-y)^{l-1}h_l(T-t)}{\sum_{l=2}^{\infty}x(1-x)^{l-1}h_l(T)}.
\label{xto0}
\end{equation}
The derivation of \eqref{xto0} is similar to that of \eqref{0to0}. 
Note from \eqref{bridgegen} 
that $X^{x,0,[0,T]}$ is a time inhomogeneous diffusion with 
time inhomogeneous infinitesimal generator
\begin{eqnarray}
\mathcal{L}_t &=& \frac{1}{2}y(1-y)\frac{\partial^2}{\partial y^2} \nonumber \\
&& \, +\, (1-y)\left[1-\frac{y\sum_{k=2}^\infty(k-1)(1-y)^{k-2}h_k(T-t)}{\sum_{k=2}^\infty(1-y)^{k-1}h_k(T-t)} \right]\frac{\partial}{\partial y}.
\label{xto0_gen}
\end{eqnarray}
The transition densities of $X^{x,0,[0,T]}$ are the same
as those of $X^{0,0,[0,T]}$, and so they 
converge as $T \to \infty$ to those of the neutral Wright-Fisher diffusion
conditioned on non-absorption.  As one would expect, the
first order coefficient in \eqref{xto0_gen} converges 
as $T \to \infty$ to $(1-2y)$, the first order coefficient in
the infinitesimal generator of the neutral Wright-Fisher diffusion
conditioned on non-absorption.

\subsection{First passage time distribution}

To determine the density of the maximum in a Wright-Fisher diffusion bridge,
 we will require the first passage time densities of
the Wright-Fisher diffusion. Let $g(\cdot;x,y)$ be the first passage time 
density from $x$ to $y$. Note that because the Wright-Fisher diffusion starting 
at $x$ may be absorbed before hitting $y$, the density $g(\cdot;x,y)$ is improper; 
that is,
\[
\int_0^\infty g(t; x, y) dt < 1.
\] 
Taking the Laplace transform of the identity
\[
f(x,y; t) = \int_0^t g(\tau;x,y)f(y,y;t-\tau) \, d\tau,
\]
we see that the Laplace transform of $g(\cdot;x,y)$ is
\begin{equation}
g^*(\lambda; x,y) = \frac{f^*(x,y;\lambda)}{f^*(y,y;\lambda)}.
\label{laplacefp}
\end{equation}
Although the Laplace transform \eqref{laplacefp} is easy to evaluate, 
it appears to be difficult to invert it explicitly
because of the denominator.

To gain more insight into first passage times, we consider moments of the 
first passage time from $x$ to $y$ conditioned on hitting $y$.
By \eqref{fp_neutral}, the first passage time distribution, conditioned on 
hitting $y$, has Laplace transform
\[
g^*(\lambda; x, y)\frac{y}{x}.
\]
Combined with \eqref{laplacefp}, the limit of this Laplace transform as 
$x \downarrow 0$ is 
\begin{equation}
\lim_{x \downarrow 0} 
\frac{f^*(x,y;\lambda)}{f^*(y,y;\lambda)} \frac{y}{x}
=
\frac{2\sum_{l=2}^\infty y(1-y)^{l-2}\phi_l(\lambda)}{f^*(y,y;\lambda)}.
\label{limitfp}
\end{equation}
It follows that 
\begin{equation}
\label{def_g_hash}
g_\#(t;y) := \lim_{x \downarrow 0} g(t;x,y) \frac{y}{x}
\end{equation}
exists and gives the density of the limit as $x \downarrow 0$
of the first passage time from $x$ to $y$ conditional
on $y$ being hit.  For later use, we record the definition
\begin{equation}
\label{def_g_diamond}
g_\diamond(t;y) := y^{-1} g_\#(t;y) = \lim_{x \downarrow 0} x^{-1} g(t;x,y).
\end{equation}

We can now use \eqref{limitfp} to calculate the mean first passage time from $0$ to $y$ conditioned on hitting $y$. The transition density satisfies the backward equation
\[
\frac{\partial}{\partial t} f(x,y;t) = \frac{1}{2}x(1-x)\frac{\partial^2}{\partial x^2}f(x,y;t).
\]
Take $y>x$, multiply by $t$, integrate from $0$ to $\infty$, and use
integration-by-parts to get
\begin{equation}
tf(x,y;t)\Big|_0^\infty - \int_0^\infty f(x,y;t) \, dt = \frac{1}{2}x(1-x)\frac{\partial ^2}{\partial x^2} \int_0^\infty t f(x,y;t) \, dt.
\label{mean_by_parts}
\end{equation}

Set
\[
\mu(x,y) := \int_0^\infty t f(x,y;t) \, dt.
\]
Use the fact that $\int_0^\infty f(x,y;t) \, dt = 2x/y$ to rewrite \eqref{mean_by_parts} as 
\[
\frac{1}{2}x(1-x)\frac{\partial^2}{\partial x^2}\mu(x,y) = -2x/y.
\]
This ordinary differential equation has the general solution
\begin{equation}
\mu(x,y) = -\frac{4}{y}(1-x)\log(1-x) + C(y)x + D(y).
\end{equation}
Differentiating \eqref{asymp} and sending $\lambda \downarrow 0$, 
we find that asymptotically as $x\downarrow 0$,
\begin{eqnarray*}
\mu(x,y) &\sim& 2x\sum_{l=2}^\infty (1-y)^{l-2}\sum_{k=l}^\infty \frac{2}{k(k-1)} \\
&=& -\frac{4x}{1-y}\log y.
\end{eqnarray*}
Thus, 
\[
\frac{4x}{y} + C(y)x + D(y) \equiv -\frac{4x}{1-y}\log y
\]
for small $x$,
and hence
\begin{equation}
\mu(x,y) = \frac{4}{y}\left[-(1-x)\log(1-x)-x\right]-4\frac{x}{1-y}\log y.
\end{equation}

To find the mean first passage time from $0$ to $y$ 
conditional on $y$ being hit (or, more correctly, the mean of the
limit as $x \downarrow 0$ of the first passage time from $x$
to $y$ conditional on $y$ being hit), differentiate \eqref{limitfp}, set $\lambda = 0$, and recall that $f^*(y,y,0) = 2$ to get
\begin{equation}
\frac{2\sum_{l=2}^\infty y(1-y)^{l-2}\sum_{k=l}^\infty \frac{2}{k(k-1)}}{2}-\frac{2\mu(y,y)}{4}
=
2 + 2\frac{1-y}{y}\log(1-y).
\label{fpcomplicated}
\end{equation}
Note that this mean increases monotonically from $0$ to $2$ 
as $y$ goes from $0$ to $1$.

\subsection{Joint density of a maximum and time to hitting in a bridge}
For the class of diffusions with inaccessible boundaries, \citet{MR891709} 
studied the joint density of a maximum and it's hitting time. This theory is not directly applicable to the Wright-Fisher diffusion because of the absorbing boundaries. However, we may condition the Wright-Fisher process to not be absorbed, thereby making the boundaries inaccessible. By an argument similar to that made in Section 2 for $h$-transforms, the bridges of this process are the same as the bridges of the unconditioned process. The transition density, $\tilde{f}(x,y;t)$ and infinitesimal generator, $\tilde{\mathcal{L}}$ of the conditioned process are given in \eqref{noabsorb_transition} and \eqref{noabsorb_generator}, respectively. We will also need the first passage time density for the conditioned process,
\[
\tilde{g}(t; x,y) = e^t x^{-1}(1-x)^{-1}g(t;x,y)y(1-y),
\]
along with its scale density,
\[
S(x) = x^{-2}(1-x)^{-2}
\]
and speed density
\[
m(x) = x(1-x).
\]
Applying the formula in Theorem A of \citet{MR891709}, we find that the joint density of the maximum and time of hitting for an arbitrary bridge from $x$ to $z$ in time $T$ is
\[
\frac{g(t;x,y)g(T-s;z,y)z^{-1}(1-z)^{-1}}{f(x,z;T)}.
\]
Taking limits as $x$, $z \downarrow 0$, we see that joint density for a bridge from $0$ to $0$ is
\[
2\frac{g_\diamond(t;y)g_\diamond(T-t;y)}{\sum_{m=2}^\infty h_m(T)}.
\]

\subsection{Maximum in a bridge}

Let $M^{x,z,[0,T]}$ be the maximum of the bridge
$\{X_t^{x,z,[0,T]}, \, 0  \le t \le T\}$, where
$0 \le x,z \le 1$.  

The occurrence of the event $\{M^{x,z,[0,T]} \geq y\}$ is equivalent
to the Wright-Fisher diffusion making a first passage from $x$
to $y$ at some time $t \in [0,T]$ and then going on to hit $z$ at time $T$.
Recalling that $g(\cdot;x,y)$ is the density of the first passage
from $x$ to $y$, for $0 < x,z < 1$ we have
\begin{equation}
\label{x_z_bridge_max}
\mathbb{P}\{M^{x,z,[0,T]}\geq y\} 
= \frac{
\int_0^T g(t;x,y)f(y,z;T-t) \, dt
}
{f(x,z;T)}.
\end{equation}

We wish to obtain an expression for $\mathbb{P}\{M^{0,0,[0,T]}\geq y\}$.
Multiply the numerator and denominator of the right-hand side
of \eqref{x_z_bridge_max} by $x^{-1}$, re-write the numerator using the
relationship
\[
f(y,x;T-t) = \frac{x^{-1}(1-x)^{-1}}{y^{-1}(1-y)^{-1}}f(x,y;T-t)
\]
that follows from the reversibility of the neutral Wright-Fisher
process with respect to the speed measure $y^{-1}(1-y)^{-1} \, dy$,
and $x,y \downarrow 0$ to get
\[
\begin{split}
& \mathbb{P}\{M^{0,0,[0,T]} \geq y ) \\
& \quad = \frac{y(1-y)
\int_0^T g_\diamond (t;y)
\sum_{i=1}^\infty (2i+1)i(i+1)P_{i-1}(1 - 2y)e^{-\frac{1}{2}i(i+1)(T-t)}
\,dt
}
{
\sum_{i=1}^\infty (2i+1)i(i+1)e^{-\frac{1}{2}i(i+1)T}
},\\
\end{split}
\]
where $g_\diamond$ was defined in \eqref{def_g_diamond} and
the sequence of polynomials $(P_n)_{n=0}^\infty$ are defined
in the Appendix.

The Laplace transform of $t \mapsto g_\#(t;y) = y g_\diamond(t;y)$
is given by \eqref{limitfp}.  Although the numerator and denominator  
of \eqref{limitfp} can be computed accurately using the orthogonal function expansion, however there is not a simple way to invert the Laplace transform of the first passage time. 

If we write the Laplace transform of $g_\#(t;y)$
\begin{equation}
g_\#^*(\lambda;y) =
\frac{
\lim_{x \downarrow 0}
\frac{1}{2}f^*(x,y; \lambda)/(x/y)}{\frac{1}{2}f^*(y,y; \lambda)},
\label{transform:0}
\end{equation}
we see that the numerator and denominator are both
Laplace transforms of probability distributions
because Green function of the neutral Wright-Fisher diffusion is
given by
\[
f^*(x,y;0) = \int_0^\infty f(x,y; t) \, dt = 2\frac{x}{y}.
\]

Equation \eqref{transform:0} can be rewritten as
\[
g_\#^*(\lambda;y) \frac{1}{2}f^*(y,y; \lambda) 
= \lim_{x \downarrow 0} \frac{1}{2}f^*(x,y; \lambda)\frac{y}{x},
\]
which implies the convolution equation
\begin{equation}
\label{convolution_equation}
g_\#(\cdot;y) \ast \left(\frac{1}{2}f(y,y; \cdot)\right)
= \lim_{x \downarrow 0} \frac{1}{2} f(x,y; \cdot)\frac{y}{x}.
\end{equation}

The easiest way to solve this equation numerically is by discretization.
Take $\epsilon > 0$ and positive integer $K$.  Let $P^{\epsilon,K}$
and $Q^{\epsilon, K}$ be the discrete probability distributions on the
set $\{0,\epsilon, 2 \epsilon, \ldots\}$ given by 
\[
a_k^{\epsilon, K} := P^{\epsilon, K}(\{k \epsilon\})
:= 
\begin{cases}
\int_0^{\epsilon/2} \lim_{x \downarrow 0} \frac{1}{2} f(x,y; t)\frac{y}{x} \, dt,
& \quad k=0, \\
\int_{(k-1/2)\epsilon}^{(k+1/2)\epsilon} 
\lim_{x \downarrow 0} \frac{1}{2} f(x,y; t)\frac{y}{x} \, dt, 
& \quad 1 \le k \le K-1,\\
\int_{(K-1/2)\epsilon}^\infty 
\lim_{x \downarrow 0} \frac{1}{2} f(x,y; t)\frac{y}{x} \, dt, 
& \quad k = K,\\
0, & \quad k >  K,
\end{cases}
\]
and
\[
b_k^{\epsilon, K} := Q^{\epsilon, K}(\{k \epsilon\})
:= 
\begin{cases}
\int_0^{\epsilon/2} \frac{1}{2}f(y,y; t) \, dt,& \quad k=0, \\
\int_{(k-1/2)\epsilon}^{(k+1/2)\epsilon} \frac{1}{2}f(y,y; t) \, dt, & \quad 1 \le k \le K-1,\\
\int_{(K-1/2)\epsilon}^\infty \frac{1}{2}f(y,y; t) \, dt, & \quad k = K,\\
0, & \quad k > K.
\end{cases}
\]
Note that the quantities $a_k^{\epsilon, K}$ and $b_k^{\epsilon, K}$
can be computed accurately using orthogonal function expansions.

Equation \eqref{convolution_equation} implies that if 
$R^{\epsilon, K}$ is the probability distribution on
the set $\{0,\epsilon, 2 \epsilon, \ldots\}$ given by
\[
R^{\epsilon, K}(\{k \epsilon\})
:= 
\begin{cases}
\int_0^{\epsilon/2} g_\#(t;y) \, dt,& \quad k=0, \\
\int_{(k-1/2)\epsilon}^{(k+1/2)\epsilon} g_\#(t;y) \, dt, & \quad 1 \le k \le K-1,\\
\int_{(K-1/2)\epsilon}^\infty g_\#(t;y) \, dt, & \quad k = K,\\
0, & k > K,
\end{cases}
\]
then $P^{\epsilon, K}$ should be approximately the convolution
$Q^{\epsilon, K} \ast R^{\epsilon, K}$.
That is, $P^{\epsilon, K}(\{k \epsilon\})$ should be approximately
$c_k$ for $0 \le k \le K$, where $c_0, \ldots, c_K$ is the solution
of the system of equations 
\[
a_k = \sum_{j=0}^k c_j b_{k-j}, \quad 0 \le k \le K.
\]
Therefore, $c_0 = a_0/b_0$ and we obtain $c_1, \ldots, c_K$ recursively by 
\begin{equation}
c_k = (a_k - \sum_{j=0}^{k-1} c_j b_{k-j}) / b_0.
\label{ccalc}
\end{equation}

Thus,
\begin{equation}
\label{max_series}
\begin{split}
&  \mathbb{P}\{M^{0,0,[0,T]} \geq y \} \\
& \quad = \frac{(1-y)
\sum_{i=1}^\infty (2i+1)i(i+1)P_{i-1}(w)g_\#^*\big (\frac{1}{2}i(i+1);T,y\big )
}
{
\sum_{i=1}^\infty (2i+1)i(i+1)e^{-\frac{1}{2}i(i+1)T}
}, \\
\end{split}
\end{equation}
where 
\begin{eqnarray*}
g_\#^*\big (\lambda;T,y\big ) &=& \int_0^T e^{-\lambda (T-t)}g_\#(t;y)dt
\nonumber \\
&\approx& \sum_{k=0}^K
\ind \big \{(k+1/2)\epsilon \leq T\big \}
\exp \Big\{
\lambda \big (T - (k+1/2)\epsilon\big )
\Big\} c(k).
\end{eqnarray*}

\subsection{Numerical calculations}
The infinite series in \eqref{max_series} was approximated using
the first $3000$ terms.  
The step size in the discrete first passage time approximation was taken to be 
$\epsilon=0.001$ 
and the number of points was taken to be $K=5000$.
\bigskip

\begin{center}
Distribution function of the maximum in a bridge $M$.
\medskip

\begin{tabular} {c|l l l l l l l l l l l l l l l l l l l l l}
\hline
$T$&0.05&0.10&0.15&0.20&0.25&0.30&0.35&0.40&0.45&0.50\\
\hline
0.5&0.0&0.02&0.17&0.43&0.66&0.83&0.92&0.96&0.99&0.99\\
1.0&&&0.0&0.02&0.09&0.21&0.36&0.52&0.66&0.77\\
1.5&&&&0.0&0.01&0.03&0.08&0.17&0.28&0.40\\
2.0&&&&&&0.0&0.02&0.04&0.09&0.17\\
\hline
$T$&0.55&0.60&0.65&0.70&0.75&0.80&0.85&0.90&0.95&1.0\\
\hline
0.5&1.0\\
1.0&0.85&0.91&0.95&0.97&0.99&0.99&1.0\\
1.5&0.52&0.63&0.73&0.82&0.88&0.93&0.96&0.99&1.0\\
2.0&0.26&0.37&0.48&0.59&0.70&0.97&0.87&0.93&0.97&1.0\\
\end{tabular}
\end{center}
\bigskip

\begin{center}

\begin{tabular} {c|l l l l l l l l l l l l l l l l l l l l l}
\hline
$T$&0.01&0.02&0.03&0.04&0.05&0.06\\
\hline
0.1&0.00&0.01&0.14&0.37&0.59&0.76\\
\hline
$T$&0.07&0.08&0.09&0.10&0.11&0.12\\
\hline
0.1&0.86&0.93&0.96&0.98&0.99&1.0\\
\end{tabular}
\end{center}
\bigskip

\noindent
The distribution function behaves as expected. If $T$ is 0.1 the maximum 
is very small, with the distribution function shown in a separate table 
with a small scale. $M$ is less than 0.06 with probability 0.76 and less 
than 0.12 with probability 1.0. If $T$ =0.5 the maximum is still small, 
but larger than when $T=0.1$, with a probability of 0.17 of being greater than 
0.3 and a probability of 1.0 of being less than 0.55. If $T=1.0,1.5,2.0$ 
the maximum is increasingly larger with respective probabilities of exceeding 
0.5 of 0.23, 0.60, 0.83 and when $T=2$ the probability of exceeding 
0.75 is 0.30. Recall that the mean to coalescence of a population to a 
single ancestor is 2 time units.

\section{Rejection sampling Wright-Fisher bridge paths}

\subsection{General framework}

When selection is incorporated into the Wright-Fisher model, there is no 
known series formula for the transition density akin to \eqref{transition} 
(but see \citet{kimura1955stochastic} and \citet{kimura1957some} for attempts
using perturbation theory, as well as \citet{Song01032012} and \citet{steinrucken2012explicit} for methods of approximating an eigenfunction expansion computationally). Therefore, analytical results for
distributions associated with the corresponding bridge
like those we obtained in the neutral case are not available. 
Instead, we develop a rejection sampling method that can  
sample paths of Wright-Fisher diffusion bridges with genic selection 
efficiently for the purpose of investigating their properties. 

Before we explain how rejection sampling can be used to sample paths
of a Wright-Fisher bridge, we first describe the analogous, but simpler, method for
sampling paths of diffusion bridges that have distributions which are
absolutely continuous with respect to that of a Brownian bridge.
Fix $x,z \in \mathbb{R}$ and $T > 0$.
Let $\mathbb{W}$ be the distribution of Brownian bridge from $x$ to $z$
over the time interval $[0,T]$, 
and let $\mathbb{P}$ be the distribution of the path of a bridge
from $x$ to $z$ over the time interval $[0,T]$ for a diffusion with infinitesimal generator
\begin{equation}
\label{G_generator}
\mathcal{G} = a(x)\frac{\partial}{\partial x} + \frac{1}{2}\frac{\partial^2}{\partial x^2}.
\end{equation}
It follows from Girsanov's theorem (see, for example,
\citet{MR1780932}) that the probability measure $\mathbb{P}$ is absolutely continuous
with respect to $\mathbb{W}$ with Radon-Nikodym derivative (that is, density) 
\begin{equation}
\frac{d\mathbb{P}}{d\mathbb{W}}(\omega) 
= 
\exp\left\{\int_0^T a(\omega_t) \, d\omega_t - \frac{1}{2} \int_0^T a^2(\omega_t) \, dt\right\}
\label{gir}
\end{equation}
for the path $\omega$, where
the first integral in \eqref{gir} is an It\^o integral
-- see  \citet{MR2187299} for the details of the  disintegration
argument that concludes this fact about Radon-Nikodym derivatives
with respect to the Brownian bridge distribution from the usual
statement of Girsanov's theorem, which is about Radon-Nikodym derivatives
with respect to the distribution of Brownian motion.
Because a Brownian bridge can be constructed using
a simple transformation of a Brownian motion (namely,
if $B$ is a standard Brownian motion, then the process
$\{x + (B_t - \frac{t}{T} B_T) + \frac{t}{T} (z-x), \, 0 \le t \le T\}$
has the distribution $\mathbb{W}$), 
it is computationally feasible 
to obtain fine-grained samples of the Brownian bridge. 
Once we have a sequence of Brownian bridge paths, 
\eqref{gir} can be used to compute a 
likelihood ratio, and a standard rejection sampling scheme can 
then be utilized
to obtain realizations of diffusion bridge paths; 
see \citet{MR2187299} for examples of extensions to this approach. 

This method is not immediately applicable to the Wright-Fisher bridge because
its infinitesimal generator is not of the form \eqref{G_generator}.  
However, it was shown on pp 119-120
of \citet{Wright1931} that if $X$ is the Wright-Fisher process
with infinitesimal generator \eqref{gen}, then the transformation
\begin{equation}
Y_t := \arccos(1-2 X_t)
\label{fisherwf}
\end{equation}
suggested in \citet{Fisher1922} produces
a diffusion process $Y$ on the state space $[0,\pi]$ 
with infinitesimal generator
\[
\mathcal{L}_Y = \frac{1}{2}(\gamma \sin(y) - \cot(y))\frac{\partial}{\partial y} 
+ \frac{1}{2}\frac{\partial^2}{\partial y^2}.
\]

Because $Y$ has absorbing boundaries at $0$ and $\pi$, sampling paths
of bridges for $Y$ by sampling Brownian bridges 
can involve extremely high rejection rates.
More specifically,
\[
\frac{1}{2}(\gamma \sin(y) - \cot(y)) \approx -\frac{1}{2y}, \quad \text{as $y \downarrow 0$},
\]
and so the likelihood ratio \eqref{gir} becomes extremely small for paths
 that spend a significant amount of time near $0$.  A similar
phenomenon occurs near $\pi$.

To overcome the difficulty near $0$, 
we develop a rejection sampling scheme where
the proposals are realizations of a process other
than the Brownian bridge.  

As a first step, consider the Wright-Fisher diffusion conditioned to be 
eventually absorbed at $1$. By the argument given in Section 2, this process
has the same bridges as the unconditional process. It follows
from \eqref{fp} and \eqref{fp_neutral} with $y = 1$ that the probability the process 
starting from $x$ is absorbed at $1$ is
\[
h(x) := \begin{cases}
\frac{1-e^{-2\gamma x}}{1-e^{-2\gamma}}, & \gamma \neq 0, \\
x, & \gamma = 0.
\end{cases}
\]
The transition densities of the conditioned process, $f_h(x,y;t)$, are related 
to the unconditional transition densities by the usual
Doob $h$-transform formula
\[
f_h(x,y;t) := h(x)^{-1}f(x,y;t)h(y).
\] 
The corresponding infinitesimal generator is
\begin{equation}
\mathcal{L}^h := \begin{cases}
\gamma x(1-x)\cot(\gamma x)\frac{\partial}{\partial x} 
+ \frac{1}{2}x(1-x)\frac{\partial^2}{\partial x^2}, 
& \gamma \neq 0, \\
(1-x)\frac{\partial}{\partial x} 
+ \frac{1}{2}x(1-x)\frac{\partial^2}{\partial x^2}, 
& \gamma = 0.
\end{cases}
\label{condwf}
\end{equation}

Applying the transformation \eqref{fisherwf} to the process
with infinitesimal generator \eqref{condwf} results in a process with 
infinitesimal generator
\begin{equation}
\label{L_Yh_generator}
\mathcal{L}_Y^h := 
\begin{cases}
\frac{1}{2}\left(\gamma\sin(y)\coth(\gamma \sin^2(y/2))-\cot(y) \right)\frac{\partial}{\partial y} + \frac{1}{2}\frac{\partial^2}{\partial y^2}, & \gamma \ne 0,\\
\frac{1}{2}\left(\sin(y)\csc^2(y/2)-\cot(y) \right)\frac{\partial}{\partial y} + \frac{1}{2}\frac{\partial^2}{\partial y^2}, & \gamma = 0.\\
\end{cases}
\end{equation}
Note that 
\begin{equation}
\label{Y_singularity}
\frac{1}{2}\left(\gamma\sin(y)\coth(\gamma \sin^2(y/2))-\cot(y) \right) 
\approx \frac{3}{2y} \quad
\text{as $y \downarrow 0$}
\end{equation}
and
\begin{equation}
\label{Y_singularity_neutral}
\frac{1}{2}\left(\sin(y)\csc^2(y/2)-\cot(y) \right)
\approx \frac{3}{2y} \quad
\text{as $y \downarrow 0$}.
\end{equation}
Moreover, if $\mathbb{Q}$ is the distribution of a bridge from $x$ to $z$
over the time interval $[0,T]$ for some diffusion with infinitesimal generator
\[
\mathcal{G} = b(x)\frac{\partial}{\partial x} + \frac{1}{2}\frac{\partial^2}{\partial x^2}
\]
and $\mathbb{P}$ is the distribution of a bridge  
from $x$ to $z$ over the time interval $[0,T]$
for the diffusion with infinitesimal generator \eqref{G_generator}, then
\[
\begin{split}
\frac{d\mathbb{P}}{d\mathbb{Q}}(\omega) 
&= \frac{d\mathbb{P}}{d\mathbb{W}}(\omega)\frac{d\mathbb{W}}{d\mathbb{Q}}(\omega) \\
&= \frac{d\mathbb{P}}{d\mathbb{W}}(\omega) \bigg/ \frac{d\mathbb{Q}}{d\mathbb{W}}(\omega) \\
&= \exp\left\{\int_0^T a(\omega_t) - b(\omega_t) \, d\omega_t 
- \frac{1}{2} \int_0^T a^2(\omega_t) -b^2(\omega_t) \, dt\right\}. \\
\end{split}
\]
This suggests that a better rejection sampling scheme for bridges of the
process $Y$ with end points close to zero will result when the proposals
come from a diffusion with an infinitesimal generator having a
first order coefficient with a singularity at zero matching the one appearing
in both \eqref{Y_singularity} and \eqref{Y_singularity_neutral}.

For such a modified scheme to be feasible, it is necessary to work
with a proposal diffusion for which 
it is easy to simulate the associated bridges.  
We now introduce such a process.
The $4$-dimensional Bessel process is the radial part of a $4$-dimensional Brownian motion. 
That is, 
if $\{B_t = (B_t^{(i)})_{i=1}^4, \, t \ge 0\}$ 
is a vector of $4$ independent one-dimensional Brownian motions, then
\[
\beta_t := |B_t| = \sqrt{(B_t^{(1)})^2+(B_t^{(2)})^2+(B_t^{(3)})^2+(B_t^{(4)})^2}, 
\quad t \ge 0,
\]
is a $4$-dimensional Bessel process (see \citet[Section XI.1]{MR1725357}
for a thorough discussion of Bessel processes).
The $4$-dimensional Bessel process is a diffusion with infinitesimal generator 
\[
\mathcal{B} := \frac{3}{2}\frac{1}{x}\frac{\partial}{\partial x} 
+ \frac{1}{2}\frac{\partial^2}{\partial x^2}.
\]

Letting $\mathbb{P}$ (resp. $\mathbb{B}$) 
be the distribution of the bridge for the 
process with infinitesimal generator \eqref{L_Yh_generator},
and hence the distribution of the transformed
Wright-Fisher diffusion $Y$, 
(resp. the $4$-dimensional Bessel
bridge) from $x$ to $z$ over the time interval $[0,T]$, we have
\begin{eqnarray}
\frac{d\mathbb{P}}{d\mathbb{B}}(\omega) 
&=& \frac{d\mathbb{P}}{d\mathbb{W}}(\omega)\frac{d\mathbb{W}}{d\mathbb{B}}(\omega) \nonumber \\
&=& \exp\left\{\int_0^T \frac{1}{2}\left(\gamma \sin(\omega_t)\coth(\alpha \sin^2(\omega_t/2)) - \cot(\omega_t)- \frac{3}{\omega_t} \right) \, d\omega_t \right. \nonumber \\
&& \left. \, - \, \frac{1}{2}\int_0^T\frac{1}{4}\left(\left(\gamma \sin(\omega_t)\coth(\alpha \sin^2(\omega_t/2)) - \cot(\omega_t)\right)^2- \frac{9}{\omega_t^2}\right) \, dt  \right\}.
\label{wfbes}
\end{eqnarray}

We next explain how to simulate a $4$-dimensional Bessel bridge.
We can construct the bridge from $u \in \mathbb{R}^4$
to $v \in \mathbb{R}^4$ over the time interval $[0,T]$
for the $4$-dimensional Brownian motion as
\[
C_t := 
\left(1 - \frac{t}{T}\right)u + \frac{t}{T} v 
+ \left(B_t - \frac{t}{T} B_T\right),
\]
where $B_0 = 0$.  The distribution of
$u + B_T$ conditional on $|u + B_T| = z$
has density proportional to $w \mapsto \exp(w \cdot u/T)$
with respect to the normalized surface measure
on the sphere centered at the origin with radius $y$,
where $w \cdot u$ is the usual scalar product of the two
vectors $w,u \in \mathbb{R}^4$.  Hence, a $4$-dimensional Bessel
bridge from $x$ to $z$ over the time interval $[0,T]$
is given by
\[
\gamma_t := 
\left|
\left(1 - \frac{t}{T}\right)u + \frac{t}{T} V 
+ \left(B_t - \frac{t}{T} B_T\right)
\right|,
\]
where $B_0=0$, $u \in \mathbb{R}^4$ is any
vector with $|u| = x$, and $V$ is random vector
taking values on the sphere centered at the origin 
with radius $z$ that is independent of $B$
and has a density with respect to the normalized surface measure
that is proportional to $w \mapsto \exp(w \cdot u/T)$.  Note
that the random vector $V/z$, which takes values
on the unit sphere centered at the origin, has
a Fisher -- von Mises distribution with 
mean vector $u/x$ and concentration
parameter $xz/T$ (see, for example,\citet[Ch. 15]{MR560319}).  

Increasing the strength of natural selection causes the
Wright-Fisher bridge to move faster for intermediate frequencies, 
but the method proposed above uses the same $4$-dimensional Bessel
bridge regardless of the value of the selection parameter $\gamma$,
and so the rejection rate can become very high for large
values of $\gamma$. To deal with this phenomenon, we
introduce the following further refinement to the proposal process.

With $\mathbb{P}$ the distribution of the transformed Wright-Fisher bridge
from $x$ to $z$ over the time interval $[0,T]$ as above, let
$\omega^\epsilon:[0,T] \to [0,\pi]$, 
$\epsilon > 0$, be the path with $\omega_0^\epsilon=x$
and $\omega_T^\epsilon = z$ that maximizes
\[
\omega \mapsto \mathbb{P}
\left\{\omega' : 
\sup_{0 \le t \le T} |\omega_t' - \omega_t| \le \epsilon
\right\}. 
\] 
Then, $\omega^\epsilon$ converges as
$\epsilon \downarrow 0$ to a path $\omega^*$.  Heuristically,
we can think of $\omega^*$ as the path that has ``maximum probability''
or is ``modal'' for $\mathbb{P}$. This path is sometimes called an Onsager-Machlup
function and it can be found by solving a certain variational
problem -- see, for example, \citet{MR1011252}.  For the 
transformed Wright-Fisher bridge, an analysis
of the variational problem shows that the maximum probability path satisfies
the second order ordinary differential equation
\begin{equation}
\ddot \omega^*
= 
\frac{\gamma^2}{8}\sin \omega^* - \frac{3}{4} \cot \omega^* \csc^2 \omega^*
\label{mpp}
\end{equation}
with boundary conditions $\omega_0^*=x$ and $\omega_T^*=z$.

With a solution to \eqref{mpp} in hand, it is possible to construct a better 
proposal distribution by linking together bridges that are ``close'' to the 
maximum probability path. First, choose a number of discretization points $N$ 
and take times $0 < t_1 < \ldots < t_N < T$. 
Then, sample independent random variables $U_1, U_2, \ldots, U_N$ with densities 
$g_1, g_2, \ldots, g_N$ to
be specified later. Put $t_0 = 0$, $t_{N+1} = T$, $U_0 = x$ and $U_{N+1} = z$. 
Build conditionally independent $4$-dimensional Bessel bridges from 
$U_i$ to $U_{i+1}$ over the time intervals $[t_i, t_{i+1}]$. 
The distribution of $U_i$ should be chosen so that $U_i$ is close to the 
maximum probability path at time $t_i$; we choose re-scaled Beta distributions 
with mode at the solution of \eqref{mpp} at time $t_i$.  More specifically,
we set $U_i = \pi X_i$, where $X_i$ has the Beta distribution with parameters
\[
\left(\frac{1+\frac{x^*_{t_i}}{\pi}(\theta - 2)}{1-\frac{x^*_{t_i}}{\pi}},
 \theta\right).
 \]
for some free parameter $\theta$.
We used the particular value $\theta = 50$ for the examples in this paper,
but other value of $\theta$ could be used in a given situation
in an attempt to optimize the frequency of rejection.

By stringing these bridges together, we get a path going from $x$ to $z$
over the time interval $[0,T]$. However, the distribution of
this path is certainly not that of the $4$-dimensional Bessel
bridge because of the manner in which we have chosen
the endpoints of the component bridges.  Therefore, we can't simply
use the Radon-Nikodym derivative \eqref{wfbes} as it stands
to construct a rejection sampling procedure.  Rather, if we
let $\mathbb{Q}$ be the distribution of the path built
by stringing the bridges together, then
we must accept a path $\omega$ with probability proportional to
\begin{equation}
\frac{d\mathbb{P}}{d\mathbb{B}}(\omega)\frac{d\mathbb{B}}{d\mathbb{Q}}(\omega).
\end{equation}
Note that
\begin{equation}
\frac{d\mathbb{B}}{d\mathbb{Q}}(\omega) = \frac{\prod_{i=0}^N \rho(\omega_{t_i},\omega_{t_{i+1}}; t_{i+1}-t_i)}{\rho(x,z;T)\prod_{i=1}^N g_i(\omega_{t_i})},
\end{equation}
where
\begin{equation}
\rho(x,z;t) := I_1\left(\frac{xy}{t}\right)\frac{y^2}{xt}e^{-\frac{x^2+z^2}{2t}}
\end{equation}
is the transition density of the $4$-dimensional Bessel process
with $I_\nu$ the modified Bessel function of the first kind. 

\begin{figure}[tp] 
   \centering
   \includegraphics[width=\textwidth]{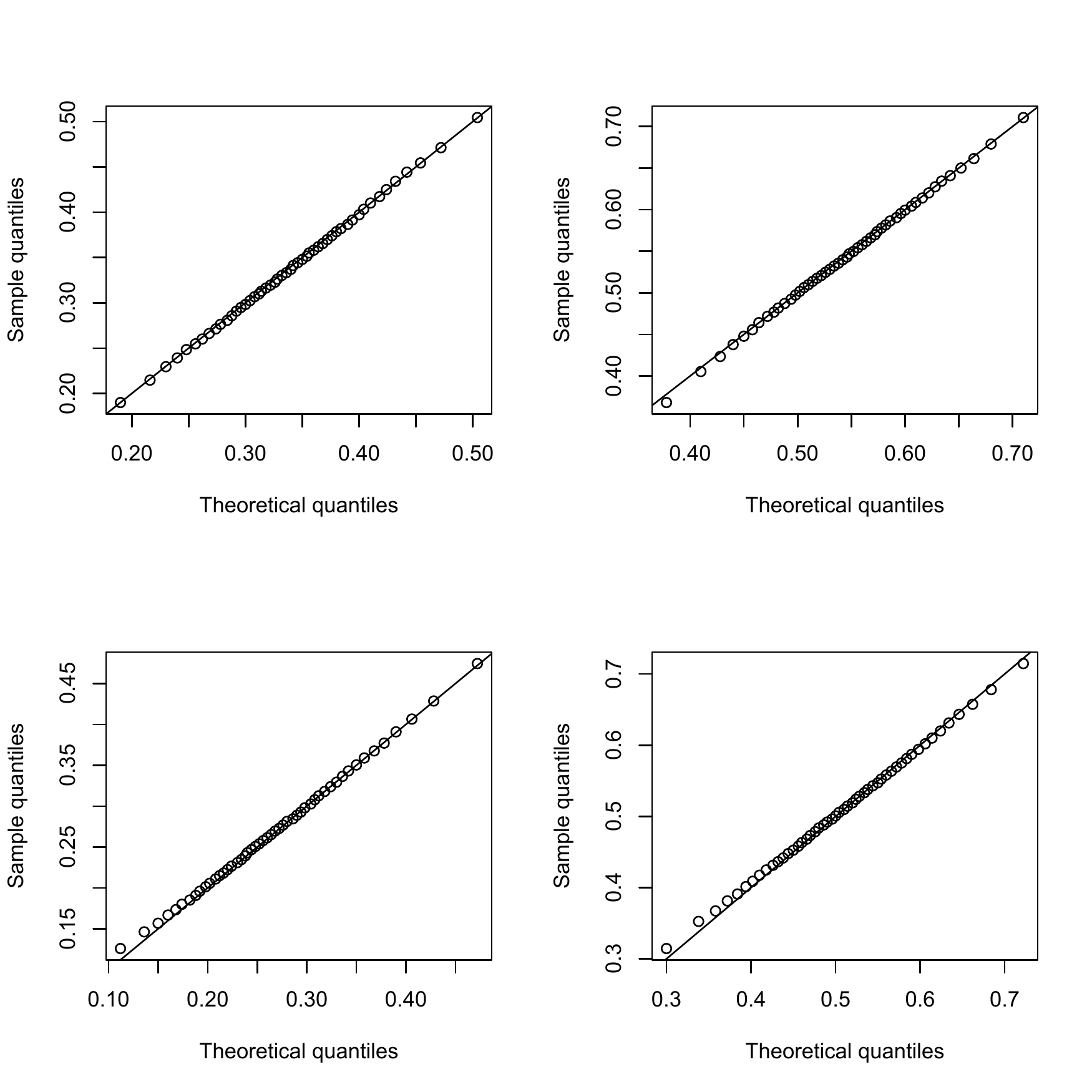} 
   \caption{Q-Q plot showing the accuracy of the rejection sampling scheme. Theoretical quantiles were calculated using the method of 
\citet{Song01032012} and sample quantiles are determined from $1000$ bridges simulated using the method described in the text.
The bridge goes from $x = 0.2$ to $z = 0.7$ over
the time interval $[0,T] = [0,0.1]$. The left panels correspond to $t = 0.03$ 
and the right panels correspond to $t = 0.07$. 
The top row corresponds to $\gamma = 10$ and the bottom row to $\gamma = 50$.}
   \label{qq_plot_6_15_12}
\end{figure}

To demonstrate the effectiveness of the rejection sampling scheme, Figure \ref{qq_plot_6_15_12} shows Q-Q plots of the one-dimensional
marginal at time $t$
of a Wright-Fisher bridge with genic selection as estimated 
using the rejection sampler compared to an approximation 
that uses the method of \citet{Song01032012} to compute the 
cumulative distribution function of the marginal. 
For both rows, the bridge goes from $x = .2$ to $z = 0.7$ over
the time interval $[0,T] = [0,0.1]$. The left panels correspond to $t = 0.03$ 
and the right panels correspond to $t = 0.07$. The top row corresponds to 
$\gamma = 10$ and the bottom row to $\gamma = 50$, 
demonstrating the effectiveness of the rejection sampling scheme over a wide range of selection coefficients.

\begin{figure}[tp] 
   \centering
   \includegraphics[width=\textwidth]{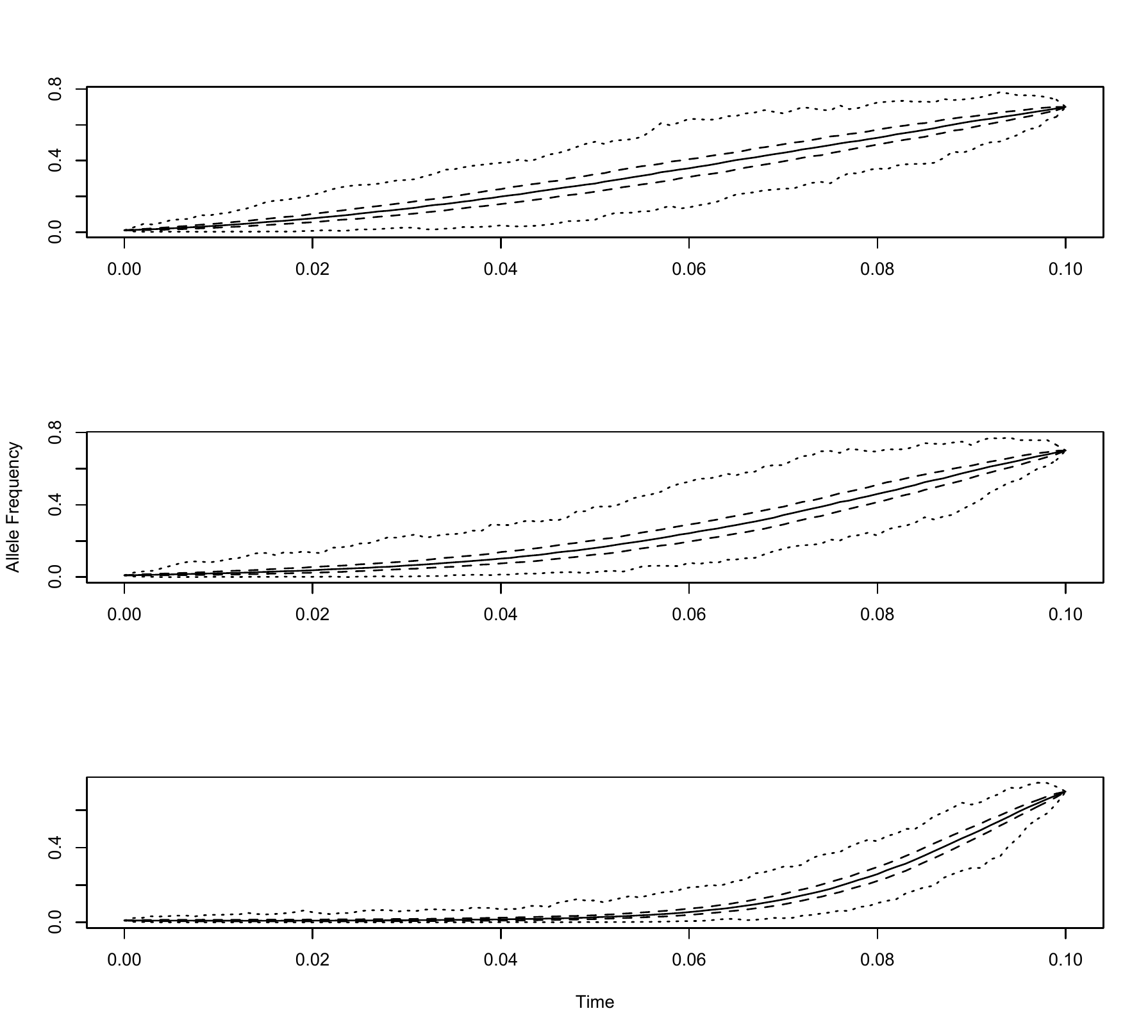} 
   \caption{Plot showing the properties of bridge paths as the strength of selection increases. 
Each bridge is from $x = 0.01$ to $z = 0.8$ over the time interval 
$[0,T] = [0,0.1]$.  The successive selection coefficients 
are $\gamma = 0$, $\gamma = 50$ and $\gamma = 100$.  
   For each selection coefficient, pointwise 0\%, 25\%, 50\%, 75\% and 100\% 
   quantiles are calculated. Solid line is the 50\% quantile, 
   dashed line indicates 25\% and 75\% quantiles, 
   and the dotted line indicates 0\% and 100\% quantiles.}
   \label{quantile}
\end{figure}

Figure \ref{quantile} demonstrates the behavior of a Wright-Fisher diffusion bridge as the selection coefficient increases. A bridge from $x = 0.01$ to $z = 0.8$ over the time interval 
$[0,T] = [0,0.1]$ is shown for $\gamma = 0$, $\gamma = 50$ and $\gamma = 100$. 
As the selection coefficient increases, the proportion of time the bridge 
spends near the boundary also increases, because the Wright-Fisher diffusion 
moves faster when it is away from the boundaries. In addition, 
the paths that the bridge takes become more tightly centered around the most 
probable path as the selection coefficient increases.

\begin{figure}[tp] 
   \centering
   \includegraphics[width=\textwidth]{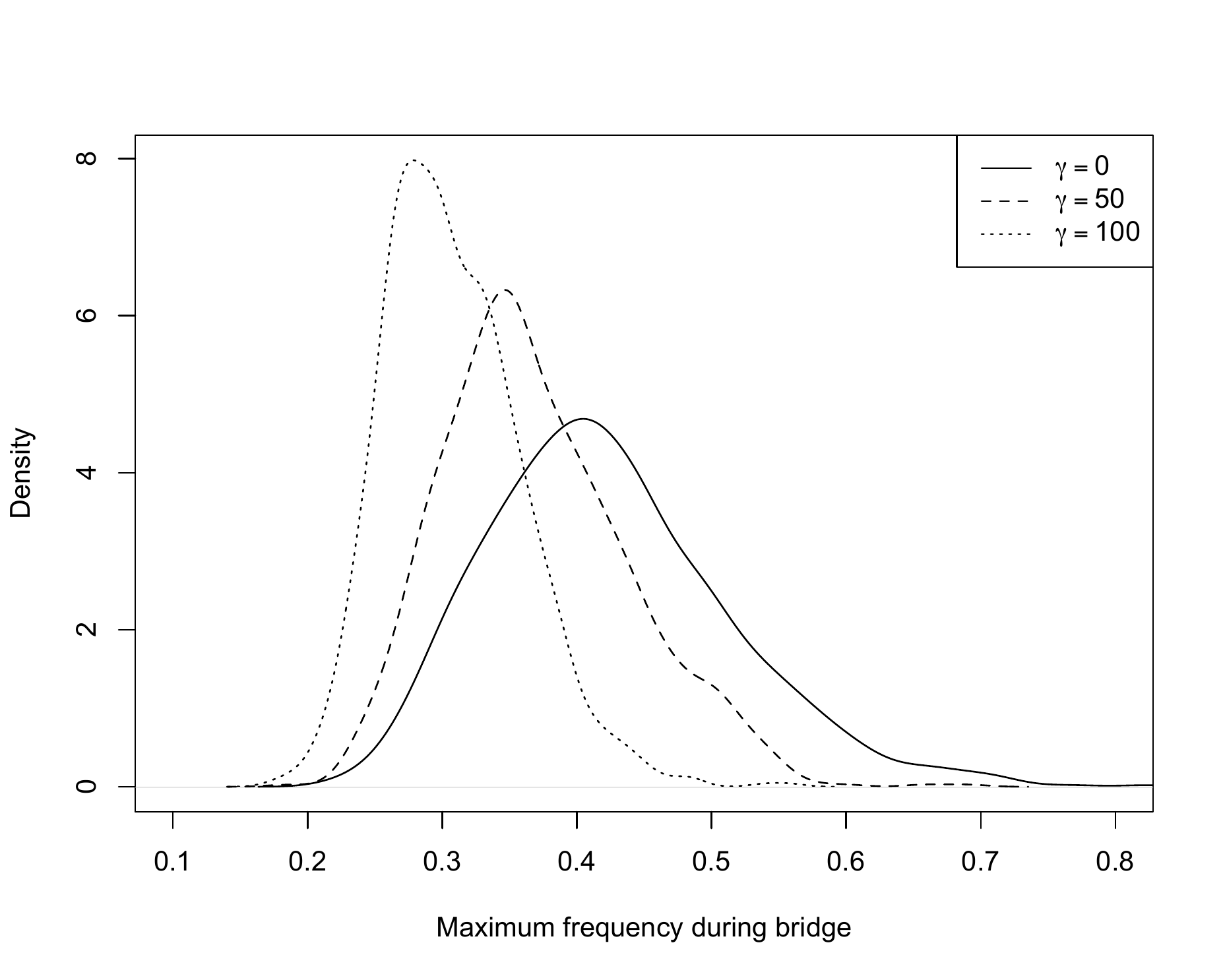} 
   \caption{Densities of the maximum in a $0$ to $0$ bridge 
   over the time interval $[0,T] = [0,0.1]$ for the selection strengths  
$\gamma = 0$, $\gamma = 50$ and $\gamma = 100$.}
   \label{max}
\end{figure}

Being able to sample Wright-Fisher bridge paths makes 
it very easy to numerically approximate the distribution and expectation of 
various functionals of the path. As an example, Figure \ref{max} shows the 
density of the maximum in a bridge from $x = 0$ to $z = 0$ over the time
interval $[0,T] = [0,0.1]$ for 
$\gamma = 0$, $\gamma = 50$ and $\gamma = 100$. Note
that the maximum in the bridge decreases as the strength of selection
increases, and also becomes more tightly concentrated around its expectation.

To gain a more quantitative understanding of the extent to which
a bridge for an allele experiencing natural selection looks different from the 
bridge for a neutral allele, it is possible to compute the 
Radon-Nikodym derivative (i.e. the likelihood ratio) of the 
distribution under selection against the distribution under neutrality. 
Using an argument similar to that which led to \eqref{wfbes}, 
the likelihood ratio is
\begin{equation}
\frac{d\mathbb{P}_\gamma}{d\mathbb{P}_0}(\omega) \propto \exp\left\{ - \frac{1}{8} \int_0^T \gamma^2 \sin^2(\omega_t) \, dt \right\},
\label{gamma}
\end{equation}
where the constant of proportionality only depends on the endpoints. A few things are immediately evident from \eqref{gamma}. First of all, the 
likelihood ratio does not depend on the sign of the selection coefficient,
 only the magnitude. This is analogous to the result
 \citet{Maruyama1974} that, 
 conditioned on eventual fixation, the sign of the selection coefficient is irrelevant to the distribution
of the Wright-Fisher diffusion path. 
Also apparent is that bridges with strong natural selection will be more likely to be found near the boundary than bridges under neutrality. Finally, because $0 \leq \sin^2(x) \leq 1$, we see that, very loosely, a bridge will look approximately neutral if
\begin{equation}
\frac{1}{8}\gamma^2 T \approx 0.
\end{equation}

\section{Discussion}

We have examined the behavior of Wright-Fisher diffusion bridges under both neutral models and models with genic selection. Although various conditioned Wright-Fisher diffusions have been studied in the past, Wright-Fisher diffusions conditioned to obtain a specific value at a predetermined time have not been studied extensively. We have elucidated some of the properties of Wright-Fisher bridges using a combination of analytical theory 
and simulations. 

In contrast to Brownian motion with drift, 
for which the distribution of a bridge does not depend 
on the magnitude of the drift coefficient,
the distribution of a Wright-Fisher bridge does depend on the magnitude of the 
selection coefficient. As one might expect, 
bridges under strong selection are more constrained than neutral bridges. 
This can clearly be seen in Figure \ref{quantile}, in which the bridge with 
$\gamma = 0$ has a broad range, but when $\gamma = 100$ the paths of the bridge 
are highly likely to be confined near the boundary at 
$0$ until quite late in the bridge. A similar conclusion can be drawn from 
Figure \ref{max} which shows the density of the maximum 
in a bridge from $0$ to $0$ over the time interval
$[0,T] = [0,0.1]$. The expected maximum of a neutral bridge is much higher 
than one with strong selection, and there is significantly more variance about 
that maximum under neutrality.

Much of the behavior of Wright-Fisher bridges under selection can be understood in terms of the likelihood ratio \eqref{gamma}. Because $\sin(x)$ takes its smallest values for $x \approx 0$ and $x \approx \pi$, very strong selection will confine a bridge of the transformed
process $Y$ to near these boundaries. Intuitively, this is because the Wright-Fisher diffusion has the largest magnitude of drift and diffusion coefficients at $x = 0.5$, and thus the diffusion moves ``faster'' when it is away from the boundaries $0$ and $1$. In order for a diffusion with a large selection coefficient to reach an interior point after a large amount of time, it must spend most of that time near the boundary.

However, these differences between selection and neutrality are mostly apparent
 in cases of extreme selection coefficients or very long times. 
 This has important implications for maximum likelihood inference of 
 selection coefficients from allele frequency time series. Because the realizations
are likely to be quite similar for a selected allele and a neutral allele
when the selection coefficient is moderate, most of the information about the selection 
coefficient comes from the end-points. Therefore, in many cases increasing 
the time-density of samples may not provide much additional information 
about the selection coefficient. Because many allelic time-series are obtained 
via costly ancient DNA techniques, this is an important consideration for the many 
researchers who are interested in the history of selection 
acting on a particular allele.

In addition to results directly concerning bridges, we have made several technical 
advances in the analysis of the Wright-Fisher diffusion.  We have developed the theory of first 
passage times of a neutral Wright-Fisher diffusion starting from low frequency 
and we were able to 
provide a closed-form for the density of the maximum in a neutral bridge that 
goes from $0$ to $0$.

While our rejection sampling scheme is similar to that of \citet{MR2187299} in some regards, there are several differences. Primarily, we do not provide exact samples, in the sense that \citet{MR2187299} does. Because we store a discrete representation of our proposal bridges in computer memory, the calculation of \eqref{wfbes} is necessarily an approximation, and hence the samples are only approximate. However, Figure \ref{qq_plot_6_15_12} shows that they are extremely accurate. 
Also, because we are concerned with a specific model, we used 
$4$-dimensional Bessel bridges, instead of Brownian bridges, in our proposal mechanism. 
This choice is superior for the Wright-Fisher diffusion because both the Bessel
 bridge and the Wright-Fisher bridge 
have boundaries at $0$ with asymptotically equivalent
singularities in the drift coefficient, while the Brownian bridge can assume 
negative values and hence result an unacceptably high rejection
rate when it is used as a proposal distribution. 
Ideally, we would sample from a proposal distribution that describes a diffusion 
that was also bounded above and had a suitable
singularity in its drift coefficient at the upper boundary; 
however, we have not yet discovered an appropriate diffusion
for which it is easy to sample the corresponding bridges. 
Finally, we make use of the ``most likely'' 
bridge path as a means of guiding samples of bridges that are likely to
be extremely different from those generated by the
$4$-dimensional Bessel bridge proposal distribution. This 
modification is akin to shifting the mean of a proposal distribution when doing 
rejection sampling of a $1$-dimensional random variable, and 
it greatly increases the efficiency of sampling. 

\section{Acknowledgments}
The authors thank M. Slatkin and B. Peter for initial discussions that led to our interest in this topic.

\bibliographystyle{plainnat}
\bibliography{bridge}

\section{Appendix}
\subsection{Eigenfunction expansions of the transition density}
Eigenfunction expansions of the Wright-Fisher transition densities 
in the case of no mutation were first explored in \citet{MR0094267}. 
The form given in \citet{Crow_Kimura1970} is
\[
f(x,y;t) = \sum_{i=1}^\infty \frac{4(2i+1)x(1-x)}{i(i+1)}C_{i-1}^{(3/2)}(1-2x)C_{i-1}^{(3/2)}(1-2y)e^{-\frac{1}{2}i(i+1)t},
\]
where $C^{(3/2)}_{i-1}$ is the Gegenbauer polynomial 
$C_{i-1}^{(\lambda)}$ with $\lambda = 3/2$.

An explicit formula for the Gegenbauer polynomial
is 
\[
C_n^{(\lambda)}(x) = \sum_{k=0}^{\lfloor n/2\rfloor} (-1)^k\frac{\Gamma(n-k+\alpha)}{\Gamma(\alpha)k!(n-2k)!}(2x)^{n-2k}
\]
The generating function for the sequence $(C_n^\lambda)_{n=0}^\infty$  is
\[
\sum_{n=0}^\infty C_n^\lambda(x) t^n = (1-2xt+t^2)^{-\lambda}.
\]
Note that  
\[
C_n^\lambda(1) =  \frac{(2\lambda)_{(n)}}{n!},
\]
and the right-hand side is $(n+1)(n+2)/2$ when $\lambda = 3/2$.

The sequence of polynomials $(C_n^{(3/2)})_{n=0}^\infty$ satisfies the
three-term recurrence
\[
n C_n^{(3/2)}(x) = (2n+1)x C_{n-1}^{(3/2)}(x) - (n+1)C_{n-2}^{(3/2)}(x)
\]
with initial conditions $C_0^{(3/2)}(x) = 1$ and $C_1^{(3/2)}(x) = 3x$. 
It is convenient in computations to use the scaled polynomials 
$P_n(x) = C_n^{(3/2)}(x)/C_n^{(3/2)}(1)$ which are bounded in modulus by unity 
on the interval $[-1,+1]$.  The corresponding three-term recurrence 
for the sequence $(P_n)_{n=0}^\infty$ is
\[
(n+2)P_n(x) = (2n+1)xP_{n-1}(x) - (n-1)P_{n-2}(x)
\]
with initial conditions $P_0(x)=1$ and $P_1(x) = x$.

The transition density written with the scaled polynomials is
\[
f(x,y;t) =x(1-x) \sum_{i=1}^\infty (2i+1)i(i+1)P_{i-1}(r)P_{i-1}(s)e^{-\frac{1}{2}i(i+1)t}.
\]
The asymptotic form of the transition density as $x \downarrow 0$ is
\begin{equation}
f(x,y;t) \approx x \sum_{i=1}^\infty (2i+1)i(i+1)P_{i-1}(s)e^{-\frac{1}{2}i(i+1)t}
\end{equation}
Also,
\[
\lim_{x,y \downarrow 0} x^{-1}f(x,y;t) = \sum_{i=1}^\infty (2i+1)i(i+1)e^{-\frac{1}{2}i(i+1)t}.
\]

We also use a form of the expansion that is formally equivalent to 
the one above -- see \citet{MR2744247}. The expansion is
\begin{equation}
f(x,y;t) = y^{-1}(1-y)^{-1}\sum_{n=2}^\infty e^{-\frac{1}{2}n(n-1)t}Q_n(x,y),
\label{eig:1}
\end{equation}
where
\begin{equation}
Q_n(x,y) := (2n-1)\sum_{m=1}^n(-1)^{n-m}\frac{m_{(n-1)}}{m!(n-m)!}\xi_m,
\end{equation}
and
\begin{equation}
\xi_m := \sum_{l=1}^{m-1}\binom{m}{l}\frac{(m-1)!}{(l-1)!(m-l-1)!}(xy)^l[(1-x)(1-y)]^{m-l}.
\end{equation}
Note that 
\[
\xi_m = x m(m-1)y(1-y)^{m-1} + O(x^2)
\]
as $x\downarrow 0$. Therefore,
\begin{equation}
f(x,y;t) \sim x \sum_{n=2}^\infty e^{-\frac{1}{2}n(n-1)t}(2n-1)\sum_{m=1}^n(-1)^{n-m}\frac{m_{(n-1)}}{m!(n-m)!}m(m-1)(1-y)^{m-2},
\label{eig:2}
\end{equation}
which is equal to \eqref{xtozero}. To calculate
\[
\lim_{x,y\downarrow 0} x^{-1}f(x,y; t) = 2\sum_{l=2}^\infty h_l(t)
\]
we observe that
\begin{equation}
\sum_{m=1}^n (-1)^{n-m}\frac{m_{(n-1)}}{m!(n-m)!}m(m-1) = n(n-1).
\end{equation}
Therefore,
\begin{equation}
2\sum_{l=2}^\infty h_l(t) = \sum_{n=2}^\infty e^{-\frac{1}{2}n(n-1)t}(2n-1)n(n-1).
\end{equation}

\newpage

\newpage

\newpage

\end{document}